\documentclass[twocolumn]{aastex631}

\usepackage{graphicx}
\usepackage[varg]{txfonts}
\usepackage{cases}
\usepackage{natbib}
\usepackage{multirow}
\usepackage{longtable}
\usepackage{booktabs}

\shorttitle{Tracking an eruptive prominence}
\shortauthors{Qingmin Zhang et al.}

\graphicspath{{./}{figures/}}

\begin{document}

\title{Tracking an eruptive intermediate prominence originating from the farside of the Sun}

\correspondingauthor{Qingmin Zhang}
\email{zhangqm@pmo.ac.cn}

\author[0000-0003-4078-2265]{Qingmin Zhang}
\affiliation{Key Laboratory of Dark Matter and Space Astronomy, Purple Mountain Observatory, Nanjing 210023, People's Republic of China}
\affiliation{State Key Laboratory of Solar Activity and Space Weather, Beijing 100190, People's Republic of China}

\author{Wenwei Pan}
\affiliation{Key Laboratory of Dark Matter and Space Astronomy, Purple Mountain Observatory, Nanjing 210023, People's Republic of China}
\affiliation{Department of Astronomy and Space Science, University of Science and Technology of China, Hefei 230026, People's Republic of China}

\author[0000-0001-8402-9748]{Beili Ying}
\affiliation{Key Laboratory of Dark Matter and Space Astronomy, Purple Mountain Observatory, Nanjing 210023, People's Republic of China}

\author[0000-0003-4655-6939]{Li Feng}
\affiliation{Key Laboratory of Dark Matter and Space Astronomy, Purple Mountain Observatory, Nanjing 210023, People's Republic of China}

\author{Yiliang Li}
\affiliation{Key Laboratory of Dark Matter and Space Astronomy, Purple Mountain Observatory, Nanjing 210023, People's Republic of China}
\affiliation{Department of Astronomy and Space Science, University of Science and Technology of China, Hefei 230026, People's Republic of China}

\author[0000-0003-2891-6267]{Xiaoli Yan}
\affiliation{Yunnan Key Laboratory of Solar physics and Space Science, Kunming 650216, People's Republic of China}
\affiliation{Yunnan Observatories, Chinese Academy of Sciences, Kunming 650216, People's Republic of China}

\author[0000-0003-0236-2243]{Liheng Yang}
\affiliation{Yunnan Key Laboratory of Solar physics and Space Science, Kunming 650216, People's Republic of China}
\affiliation{Yunnan Observatories, Chinese Academy of Sciences, Kunming 650216, People's Republic of China}

\author[0000-0002-1190-0173]{Ye Qiu}
\affiliation{Institute of Science and Technology for Deep Space Exploration, Suzhou Campus, Nanjing University, Suzhou 215163, People's Republic of China}

\author[0000-0003-3060-0480]{Jun Chen}
\affiliation{Key Laboratory of Dark Matter and Space Astronomy, Purple Mountain Observatory, Nanjing 210023, People's Republic of China}

\author[0000-0002-5431-6065]{Suli Ma}
\affiliation{State Key Laboratory of Solar Activity and Space Weather, Beijing 100190, People's Republic of China}

\begin{abstract}
In this paper, we carry out multiwavelength and multiview observations of the eruption of an intermediate prominence 
originating from the farside of the Sun on 2023 March 12.
The southeast footpoint of the prominence is located in active region (AR) 13252. 
The eruption generates a B7.8 class flare and a partial halo coronal mass ejection (CME).
The prominence takes off at 02:00 UT and accelerates for nearly three hours. 
Rotation of the southeast leg of the prominence in the counterclockwise direction is revealed by spectroscopic and imaging observations.
The apex of the prominence changes from a smooth loop to a cusp structure during the rising motion
 and the northwest leg displays a drift motion after 04:30 UT, implying a writhing motion.
Hence, the prominence eruption is most likely triggered by ideal kink instability.
For the first time, we apply the Graduated Cylindrical Shell (GCS) modeling in three-dimensional reconstruction and tracking of the prominence for nearly two hours.
Both the source region (110$\degr$E, 43$\degr$N) and northwest footpoint (162$\degr$E, 44$\degr$N) are located.
The edge-on and face-on angular widths of the prominence are $\sim$6$\degr$ and $\sim$86$\degr$, respectively.
The axis has a tilt angle of $\sim$70$\degr$ with the meridian.
The heliocentric distance of the prominence leading edge increases from $\sim$1.26\,$R_{\sun}$ to $\sim$2.27\,$R_{\sun}$.
The true speed of the CME increases from $\sim$610 to $\sim$849 km s$^{-1}$.
\end{abstract}

\keywords{Sun: prominences --- Sun: flares --- Sun: coronal mass ejections (CMEs)}

\section{Introduction} \label{intro}
Prominences are cool and dense plasmas hanging up in the hot solar corona \citep{tan95,mac10,gib18}. 
The materials are found to come from direct injections from the chromosphere \citep{wang19}, levitation \citep{zhao17}, 
or coronal condensation as a result of thermal instability \citep{yosh25,zhou25} along polarity inversion lines (PILs) \citep{mar98,chen14}.
After formation, they are supported by magnetic dips in sheared arcades \citep{au02} or magnetic flux ropes \citep[MFRs;][]{zhou18}.
Prominences are rich in dynamics, including counter streamings \citep{zir98,chen14}, magneto-thermal convection \citep{ber11}, rotations \citep{yan14b}, tornadoes \citep{wed12},
small-amplitude oscillations \citep{oka07,ning09,song24}, 
large-amplitude oscillations \citep{iso06,luna18,zqm12,zqm24a}, mass drainage \citep{jen18,dai21}, and eruptions \citep{gil00,ji03,sch13,zqm24b}. 
The triggering mechanisms of prominence eruptions include ideal helical kink instability \citep{hood81,kli04,tor04,tor05,fan05,gre07}, 
torus instability \citep{kli06,au10,xing24}, 
magnetic flux emergence \citep{chen00,chen24b}, catastrophic loss of equilibrium \citep{for91,lin00}, tether-cutting model \citep{mo01}, and breakout model \citep{ant99}.
Occasionally, a primary prominence eruption is capable of inducing another one nearby or at a remote site, which are called sympathetic eruptions \citep{tor11,shen12,li25b}.
Small-scale and large-scale prominence eruptions are likely to generate coronal jets \citep{ste15,zqm16,wy18}, flares \citep{jan15}, and coronal mass ejections \citep[CMEs;][]{for06,gre18}.
These explosions could be regularly detected in H$\alpha$ \citep{chen08,chen25,li25a}, Lyman-alpha (Ly$\alpha$) \citep{wei24,li25b}, 
extreme-ultraviolet (EUV) \citep{mie22}, and radio \citep{bas01,hua19,hou25} wavelengths.

The three-dimensional (3D) morphology and evolution of a prominence or a CME are unclear from single-view observations. 
The separate vantage points of twin spacecrafts of the Solar TErrestrial RElations Observatory \citep[STEREO;][]{kai08} in combination with the vantage point of Earth 
allow for precise 3D reconstructions and continuous tracking of prominences and CMEs.
In the past two decades, several methods and models have been developed \citep[see][and references therein]{mie10}.
The most popular one is tie-pointing (TP) or triangulation method \citep{in06,bem09,lie09,jos11,li11,pan11,tho11,tho12,shen12,bi13,how15,guo19,sah23,chen25}.
The most common feature reconstructed by TP is the leading edge of a CME and more compact structures \citep{mie08}.
At least two viewpoints are required in this technique. Correct identification of the same feature in these images is very important.
The limitation of this method might be that features at the farside of the Sun observed from either viewpoint could not be reconstructed, 
since it is impossible to find a correspondence between pixels in the images taken by these spacecrafts.

To investigate the geometry and kinematics of non-radial prominence eruptions and CMEs,
\citet{zqm21} proposed a revised cone model, which is characterized by four parameters: the length and angular width of the cone, the deflection angles in longitude and latitude directions, respectively. The tip of the cone is on the solar surface rather than at the Sun center.
Using multiwavelength and multiview observations, 
the model is successfully applied to the 3D reconstructions and tracking of bright fronts of CMEs \citep{zqm22,dai23} and the leading edges of prominences \citep{zqm24b} as far as $\sim$10\,$R_{\sun}$.
Deflection, acceleration, and expansion are perfectly uncovered based on the modeling.
The revised cone model is obviously inappropriate for reconstructions of huge prominences whose footpoints are remarkably separated.

Considering that a large number of CMEs are MFRs in nature \citep{lep90,bo98,vour13,pal18}, 
plenty of models featuring a MFR have been developed to perform 3D reconstructions of CMEs \citep[e.g.,][]{the06,wood09,is16,mo18}.
The Graduated Cylindrical Shell \citep[GCS;][]{the06,the09,the11} model is composed of two conical legs and a middle circulus connecting the two legs, resembling a croissant. 
The model has been widely applied to 3D reconstructions of CMEs observed by white-light (WL) coronagraphs, 
such as COR1 and COR2 on board the STEREO spacecrafts, C2 and C3 of the Large Angle Spectroscopic Coronagraph \citep[LASCO;][]{bru95} on board the Solar and Heliospheric Observatory (SOHO) mission,
and Metis \citep{ant20} on board the Solar Orbiter \citep[SolO;][]{mu20}. 
The geometrical and kinematical evolutions of CMEs are exhaustively investigated \citep{cx13,col13,kw14,and21,bem22,zhou23,sha24,te24,zj24,sah25}.
It should be noted that the GCS is just a geometrical model without taking into account the magnetic field,
which is different from the FRi3D model \citep{is16} and 3DCORE model \citep{mo18}.

To explore the evolutions of flux ropes or prominences originating from ARs, whose footpoints are close to each other,
\citet{zqm23} made a slight modification to the GCS model 
and applied it to the 3D reconstruction of an eruptive prominence originating from AR 13110 on 2022 September 23.
It is found that the prominence experiences a southward deflection by $\sim$15$\degr$. 
Meanwhile, the velocity increases from $\sim$246 to $\sim$708 km s$^{-1}$ during the impulsive phase of the related M1.7 class flare.
The revised GCS model has the same limitation as the revised cone model.

In this paper, we study a long, intermediate prominence, with one footpoint locating in active region (AR) 13252 on the visible side, 
while the other footpoint being on the farside as observed from Earth. Therefore, the two footpoints are considerably separated.
Both the revised cone model and revised GCS model are unsuitable for modeling.
For the first time, we apply the standard GCS model \citep{the06} to the 3D reconstruction and tracking of the prominence 
assuming a radial propagation without deflections.
The paper is organized as follows. We describe multiwavelength observations in Section~\ref{data}.
Data analysis and results of reconstructions are presented in Section~\ref{res}. 
Discussions and a brief conclusion are given in Sections~\ref{dis} and \ref{con}, respectively.

\section{Observations} \label{data}
On 2023 March 12, an intermediate prominence erupted close to the eastern limb, generating a weak B7.8 class flare and a wide CME.
The prominence eruption was detected by multiple instruments from diverse vantage points, including
the Atmospheric Imaging Assembly \citep[AIA;][]{lem12} on board the Solar Dynamics Observatory \citep[SDO;][]{pes12} spacecraft, 
the Solar Ultraviolet Imager \citep[SUVI;][]{tad19,dar22} on board the GOES-16 spacecraft, 
the Full Sun Imager (FSI) of the Extreme Ultraviolet Imager \citep[EUI;][]{ro20} on board SolO,
the H$\alpha$ Imaging Spectrograph \citep[HIS;][]{qiu22} on board the Chinese H$\alpha$ Solar Explorer \citep[CHASE;][]{li22}, 
and the Solar Corona Imager (SCI\_UV) of the Lyman-alpha (Ly$\alpha$) Solar Telescope (LST; \citealt{feng19,li19,chen24a}) 
on board the Advanced Space-based Solar Observatory (ASO-S; \citealt{gan19,gan23}).

\begin{figure}
\includegraphics[width=0.45\textwidth,clip=]{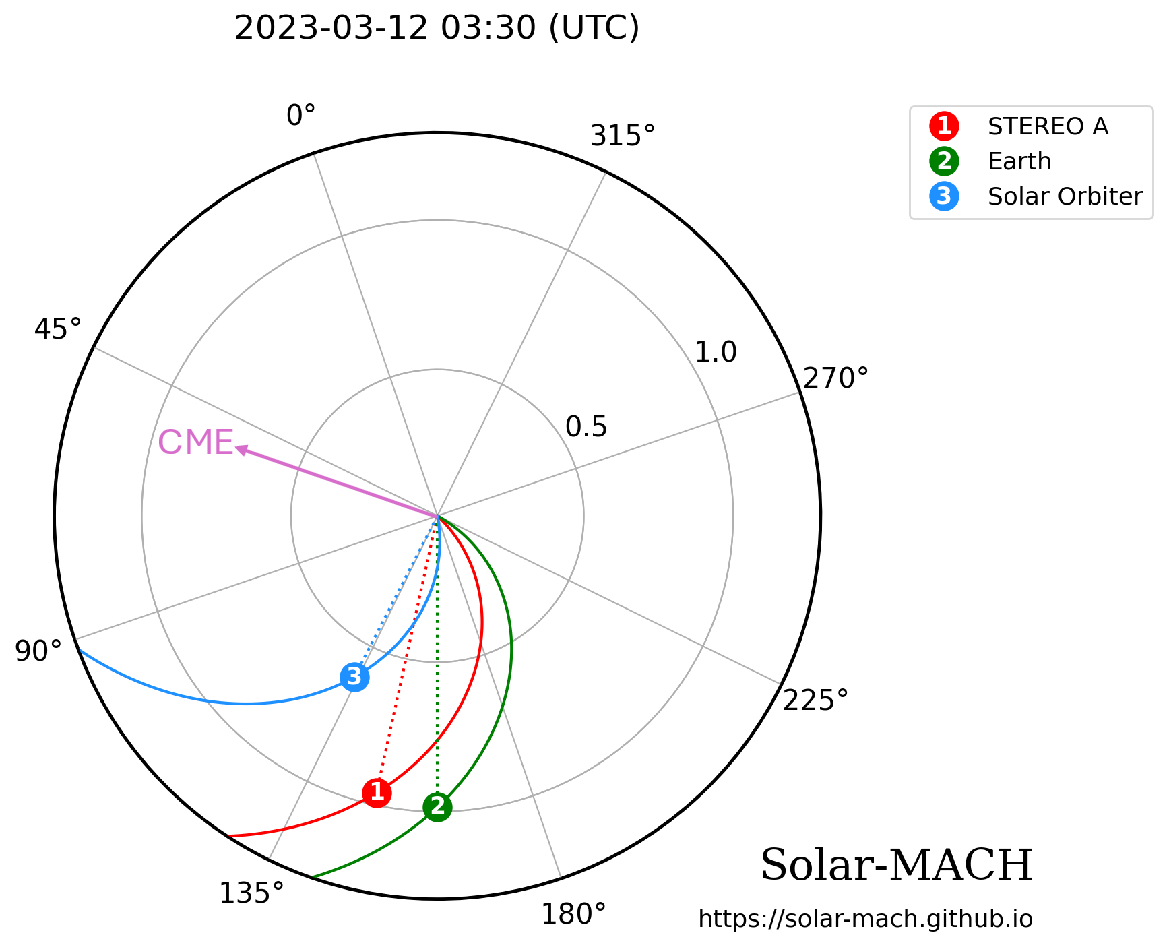}
\centering
\caption{The Solar-MACH plot at 03:30 UT on 2023 March 12, 
illustrating the locations and connectivity to the Sun of STA (red circle), SolO (blue circle), and Earth (green circle).
The purple arrow indicates the longitudinal direction of partial halo CME. 
The spiral arms are built by using the default solar wind value (400 km s$^{-1}$) on Solar-Mach.}
\label{fig1}
\end{figure}

In Figure~\ref{fig1}, the Solar-MACH plot\footnote{https://solar-mach.streamlit.app/?embedded=true} \citep{gie23} illustrates the locations of STA (red circle), SolO (blue circle), and Earth (green circle), respectively.
The properties of these instruments are summarized in Table~\ref{tab-1}, including the wavelengths, pixel sizes, time cadences, 
heliocentric distances, longitudinal and latitudinal separation angles ($\phi_0$ and $\theta_0$) with the Sun-Earth line.
SDO/AIA takes full-disk images in 7 EUV (94, 131, 171, 193, 211, 304, and 335 {\AA}) and 2 UV (1600 and 1700 {\AA}) wavelengths.
The full-disk line-of-sight (LOS) magnetograms of the photosphere are observed by the Helioseismic and Magnetic Imager \citep[HMI;][]{sch12} on board SDO.
The level\_1 data of AIA and HMI are calibrated using the \texttt{aia\_prep.pro} and \texttt{hmi\_prep.pro} in Solar Software (SSW).
The level\_2 data of EUI/FSI are processed using the \texttt{eui\_readfits.pro} and rotated to align with the solar north.
The SUVI images are also rotated and shifted slightly to align with the AIA images \citep{zqm24b}.
The ASO-S/SCI\_UV Ly$\alpha$ images are primarily processed through the correction of flat fields and dark currents, along with the normalization of exposure times. 
The daily-minimum backgrounds are subtracted to suppress stray light signals.
Then, the images are rotated, scaled, and aligned with the AIA 304 {\AA} images \citep{li25b}.
Transforms between the coordinate systems from various vantage points are described in detail \citep{zqm24b}.
Assuming that the prominence plasmas act as a luminous block with a constant source function, 
the Doppler shift of the prominence can be derived by implementing the modified cloud model which neglects background emission \citep{qiu24} 
on the high-spectral-resolution observations provided by CHASE/HIS \citep{qiu22}.
WL images of the associated CME are observed by SOHO/LASCO and COR2 of the ahead STEREO (hereafter STA).
Soft X-ray (SXR) fluxes of the B-class flare in 1$-$8 {\AA} are recorded by the GOES-16 spacecraft.

\begin{deluxetable}{ccccccc}
\tablecaption{Wavelengths, pixel sizes, cadences, heliocentric distances, 
longitudinal and latitudinal separation angles with the Sun-Earth line of the instruments on 2023 March 12. \label{tab-1}}
\tablecolumns{7}
\tablenum{1}
\tablewidth{0pt}
\tablehead{
\colhead{Instrument} &
\colhead{$\lambda$} &
\colhead{Pix. Size} &
\colhead{Cadence} &
\colhead{Dist.} &
\colhead{$\phi_0$} &
\colhead{$\theta_0$}\\
\colhead{} &
\colhead{($\AA$)} &
\colhead{(arcsec)} &
\colhead{(s)} &
\colhead{(au)} &
\colhead{(deg)} &
\colhead{(deg)}
}
\startdata
CHASE/HIS & 6562.8 & 1.0 & 60 & 1.0 & 0 & 0 \\
ASO-S/SCI\_UV & 1216 & 2.45 & 60 & 1.0 & 0 & 0 \\
SDO/AIA  & 304 & 0.6 & 12, 24 & 1.0 & 0 & 0 \\
SDO/HMI & 6173 & 0.6 & 45 & 1.0 & 0 & 0 \\
GOES-16/SUVI & 304 & 2.5 & 100 & 1.0 & 0 & 0 \\
GOES-16 & 1$-$8 & $-$ & 1 & 1.0 & 0 & 0 \\
SolO/EUI & 304 & 4.4 & 450 & 0.62 & -27.3 & 2.4 \\
STA/COR2 & WL & 15.0 & 900 & 0.97 & -12.4 & 0 \\
LASCO-C2 & WL & 11.4 & 720 & 0.99 & 0 & 0 \\
LASCO-C3 & WL & 56.0 & 720 & 0.99 & 0 & 0 \\
\enddata
\end{deluxetable}

\section{Data Analysis and Results} \label{res}
\subsection{Prominence eruption, flare, and CME} \label{cme}
\begin{figure}
\includegraphics[width=0.45\textwidth,clip=]{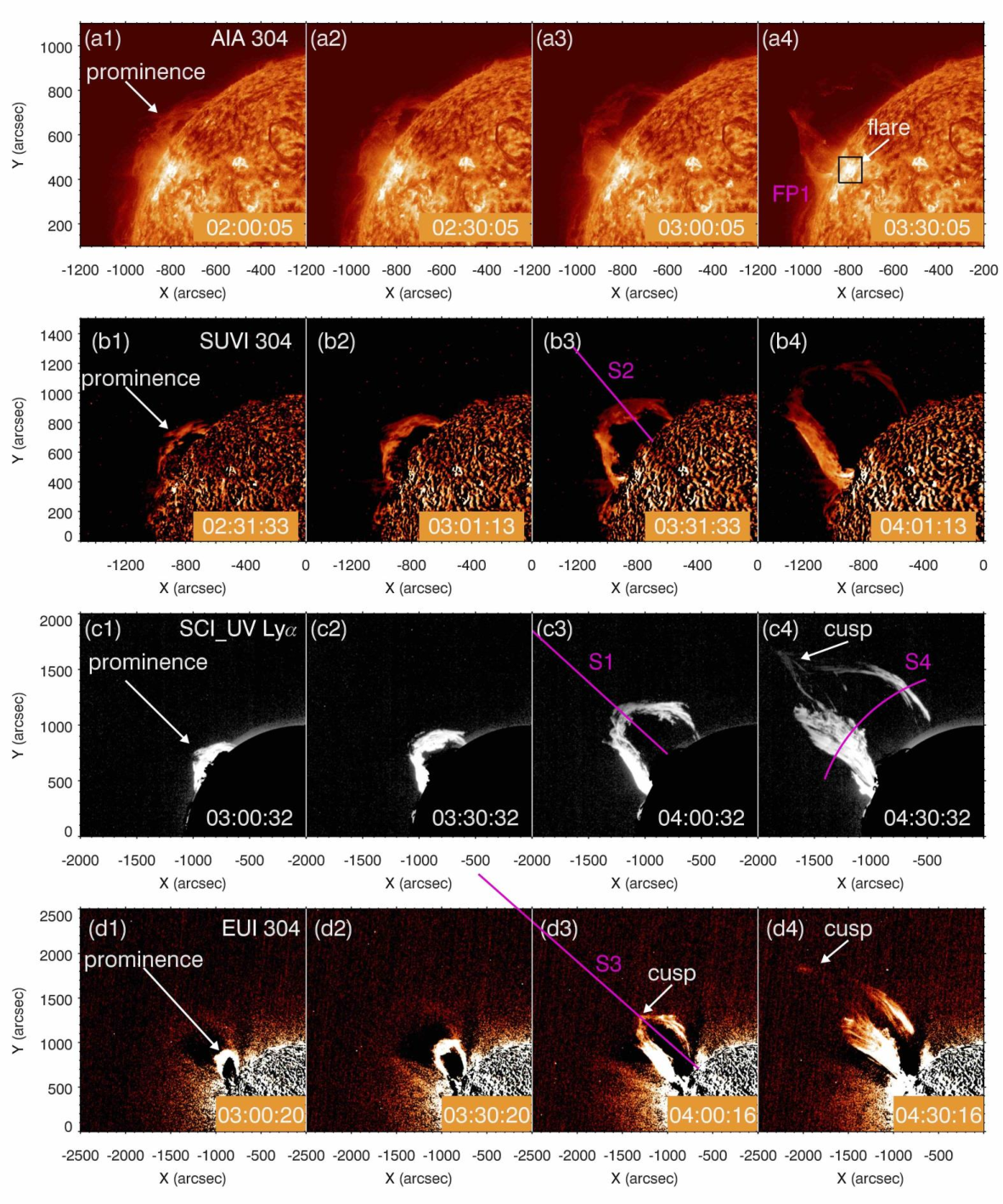}
\centering
\caption{The eruptive prominence observed by AIA 304 {\AA} (a1-a4), SUVI 304 {\AA} (b1-b4), SCI\_UV Ly$\alpha$ (c1-c4), and EUI 304 {\AA} (d1-d4), respectively.
The white arrows in the left panels point to the prominence.
In panel (a4), the black box is used to derive the EUV light curve of the flare.
In panels (c4), (d3), (d4), the white arrows point to the cusp structure.
The magenta slices in panels (b3), (c3), (d3) are used to investigate the height evolution of the prominence above the limb.}
\label{fig2}
\end{figure}

\begin{figure}
\includegraphics[width=0.40\textwidth,clip=]{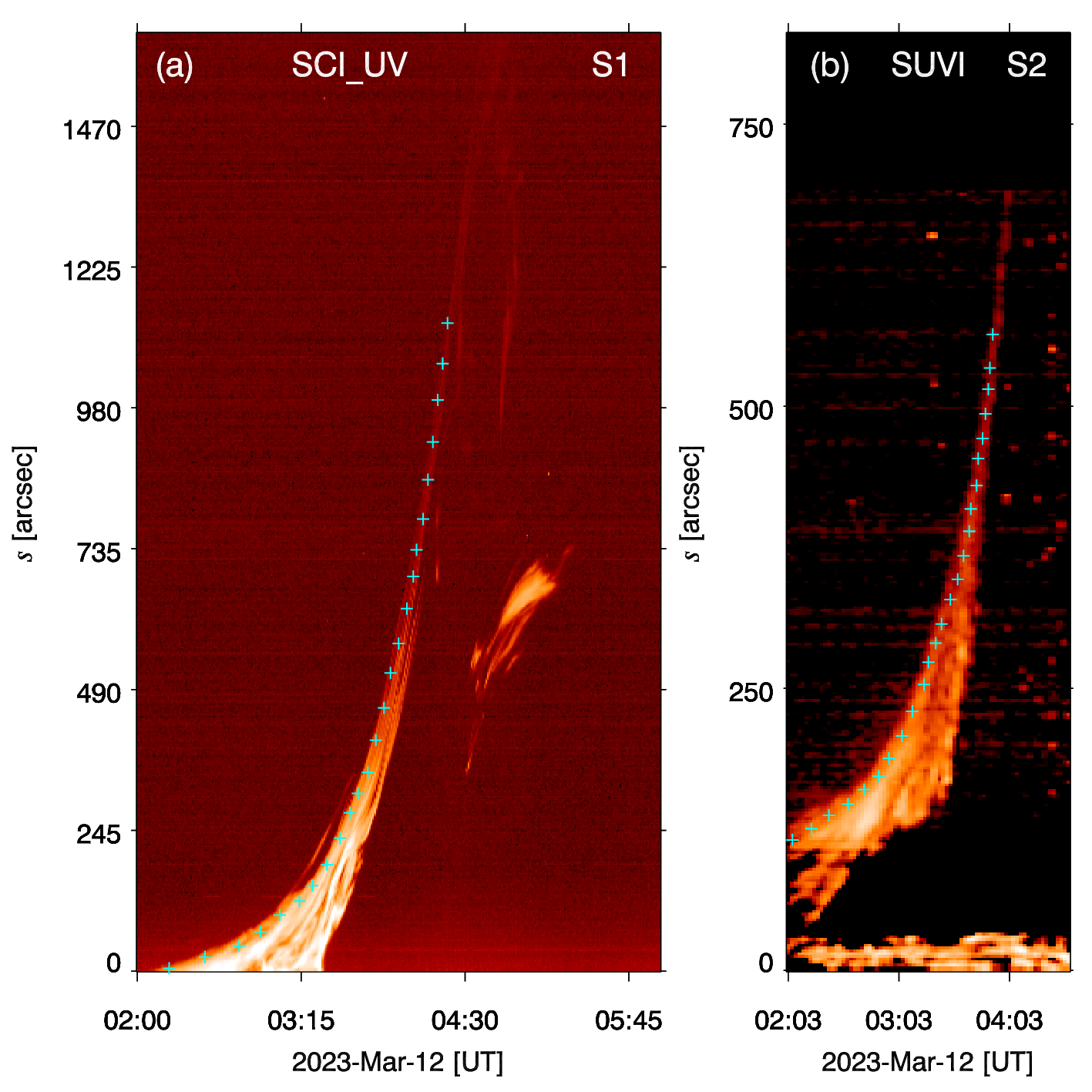}
\centering
\caption{Time-distance diagrams of S1 in SCI\_UV Ly$\alpha$ (left panel) and S2 in SUVI 304 {\AA} (right panel). 
$s=0\arcsec$ denotes the southwest endpoints of the slices.
The cyan plus symbols denote the corresponding trajectories of the prominence in the plane of the sky.}
\label{fig3}
\end{figure}

In Figure~\ref{fig2}, the eruptive prominence is displayed across four rows, from top to bottom, as observed by AIA, SUVI, SCI\_UV, and EUI, respectively.
The long, loop-like prominence starts to lift off slowly at $\sim$02:00 UT. 
The southeast footpoint (FP1) resides in AR 13252 (E58N42) close to the eastern limb, while the northwest footpoint (FP2) is behind the visible solar disk (panel (a4)).
As the prominence ascends, its apex stands out and the shape changes from a smooth loop to a cusp structure, 
as imaged by ASO-S/SCI\_UV and EUI/FSI thanks to their larger field of views (FOVs) (see panels (c4), (d3), and (d4) in Figure~\ref{fig2}).
To investigate the height evolution of the prominence, two slices (S1 in panel (c3) and S2 in panel (b3)) with widths of 5.4 Mm 
are selected and drawn with magenta lines.
Time-slice diagrams of S1 and S2 are plotted in the left and right panels of Figure~\ref{fig3}.
The corresponding trajectories of the prominence in the plane of the sky are marked with cyan plus symbols, indicating obvious acceleration during the propagation.

\begin{figure}
\includegraphics[width=0.45\textwidth,clip=]{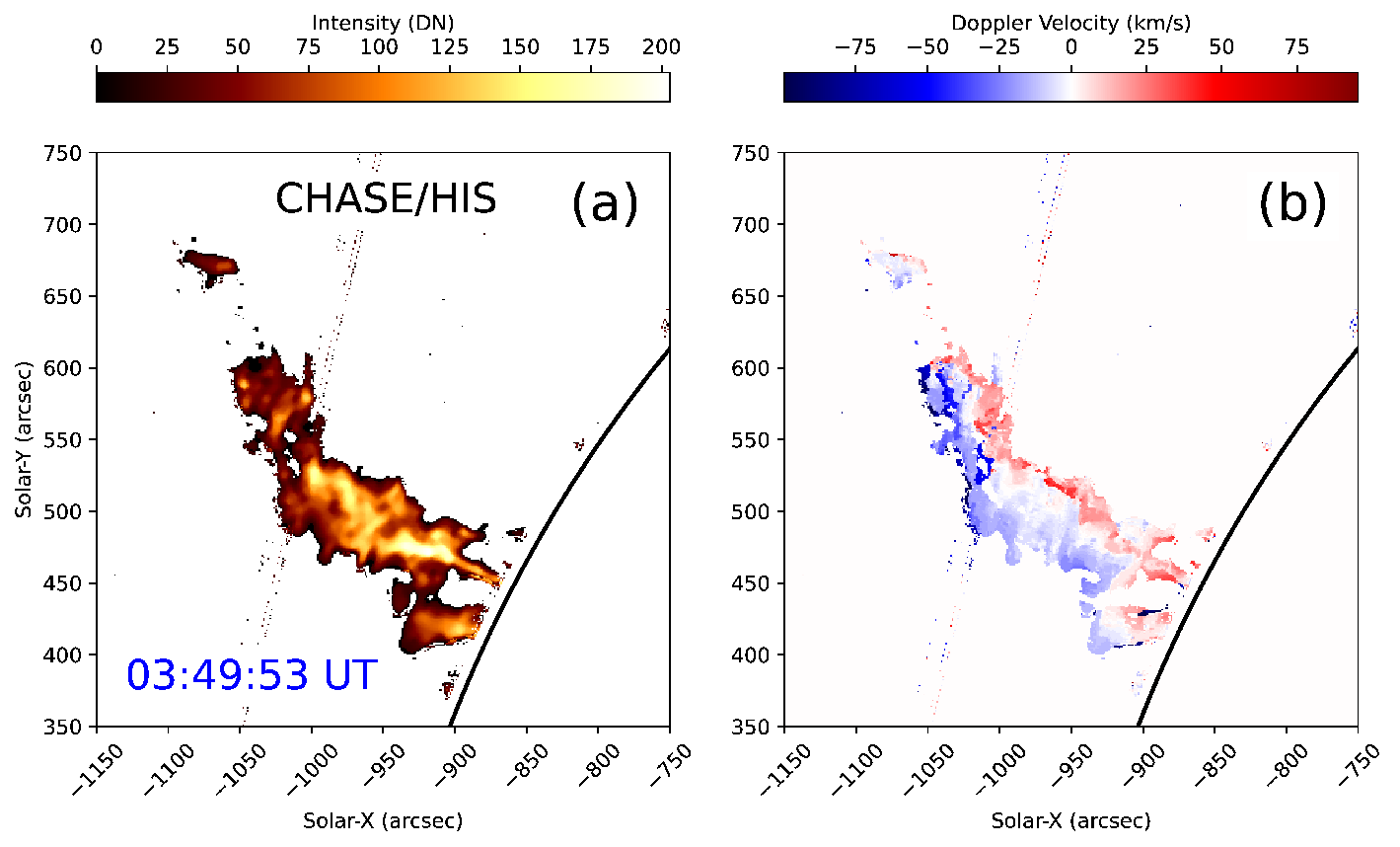}
\centering
\caption{H$\alpha$ intensity map (a) and Dopplergram (b) obtained by CHASE/HIS at 03:49:53 UT, 
showing the southeast leg of the prominence. The solar limb is drawn with a thick black line.}
\label{fig4}
\end{figure}

In Figure~\ref{fig4}, the H$\alpha$ intensity map and Dopplergram of the southeast leg of the prominence obtained by CHASE/HIS at 03:49:53 UT are displayed in the left and right panels, respectively.
Adjacent blueshifts and redshifts with velocities up to 50 km s$^{-1}$ are simultaneously observed, indicating clear rotation of the prominence along the axis \citep{liu07}.
The direction of rotation is counterclockwise viewed from above.
\citet{zhou23} studied a prominence eruption on 2013 May 13. Similarly, one of the legs undergoes rapid rotation in the counterclockwise direction during the eruption observed by SDO/AIA. 
Using the time-distance diagram of a short slice across the rotating leg, the total twist of the leg is estimated to be $\sim$7$\pi$ (see their Fig. 5).
The value is greater than the threshold of kink instability (2$\pi$-3.5$\pi$), 
which depends on the details of the MFR equilibrium, including the radial profile of the magnetic field, plasma $\beta$, and flux rope aspect ratio \citep{hood79,zz21}.
\citet{qiu24} derived the 3D velocity map of an eruptive filament and found counterclockwise rotation of the material in the filament on 2022 August 17.
\citet{yan14b} found that the filament in AR 11082 exhibits a quick uplift accompanying a counterclockwise rotation of the filament body with a total twist of $\ge$5$\pi$.
The apparent width of the spinning filament increases with time.
Using high-resolution H$\alpha$ data from the New Vacuum Solar Telescope \citep[NVST;][]{liu14}, 
\citet{yan20} studied the untwisting motion of a filament on 2019 May 7.
Adjacent redshifts and blueshifts along the filament body are clearly revealed in the Dopplergrams.
A clear helix with 3$-$4 turns (corresponding twist 6$\pi$-8$\pi$) is found in an eruptive prominence on 2011 February 24, 
which is believed to be a strong evidence of helical kink instability \citep{kum12}.

\begin{figure}
\includegraphics[width=0.45\textwidth,clip=]{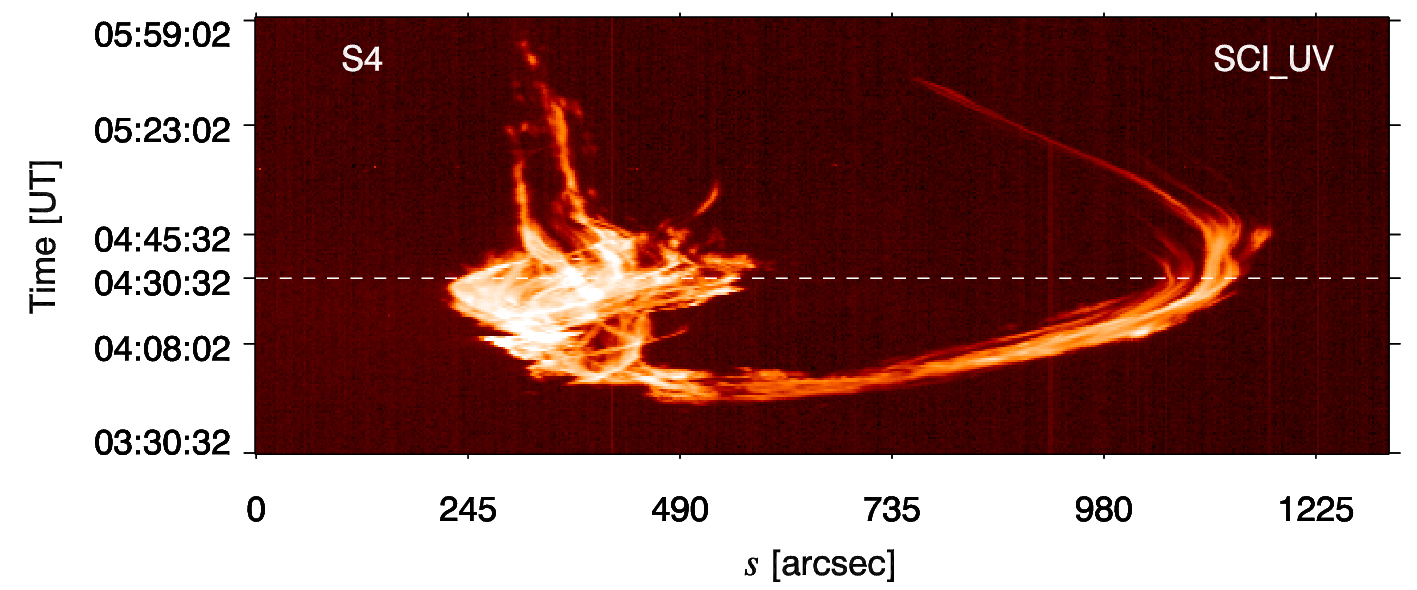}
\centering
\caption{Time-distance diagram of S4 in SCI\_UV Ly$\alpha$. $s=0\arcsec$ and $s=1309\arcsec$ denote the southeast and northwest endpoints of S4.
The horizontal dashed line denotes 04:30:32 UT, when the northwest leg starts to drift eastward.}
\label{fig5}
\end{figure}

Rotation motion of the prominence leg is also observed by SCI\_UV. In Figure~\ref{fig2}(c4), a curved slice (S4) with a width of $\sim$5.29 Mm and 
a height of $\sim$0.55\,$R_{\sun}$ above the solar limb,
is selected to investigate the rotation in Ly$\alpha$. Time-slice diagram of S4 is plotted in Figure~\ref{fig5}.
It is evident that as the prominence rotates, the width of the southeast leg increases significantly and reaches the maximum ($\sim$243.4 Mm) around 04:30:32 UT (horizontal dashed line).
Meanwhile, the northwest leg starts to drift eastward and the cusp structure becomes sharper.
Although the rotating motion is very clear in the SCI\_UV images, the total twist could not be obtained from time-distance diagram of S4, 
since the leg is composed of many fine threads. It is very difficult to track the threads to estimate the total twist.

\begin{figure}
\includegraphics[width=0.45\textwidth,clip=]{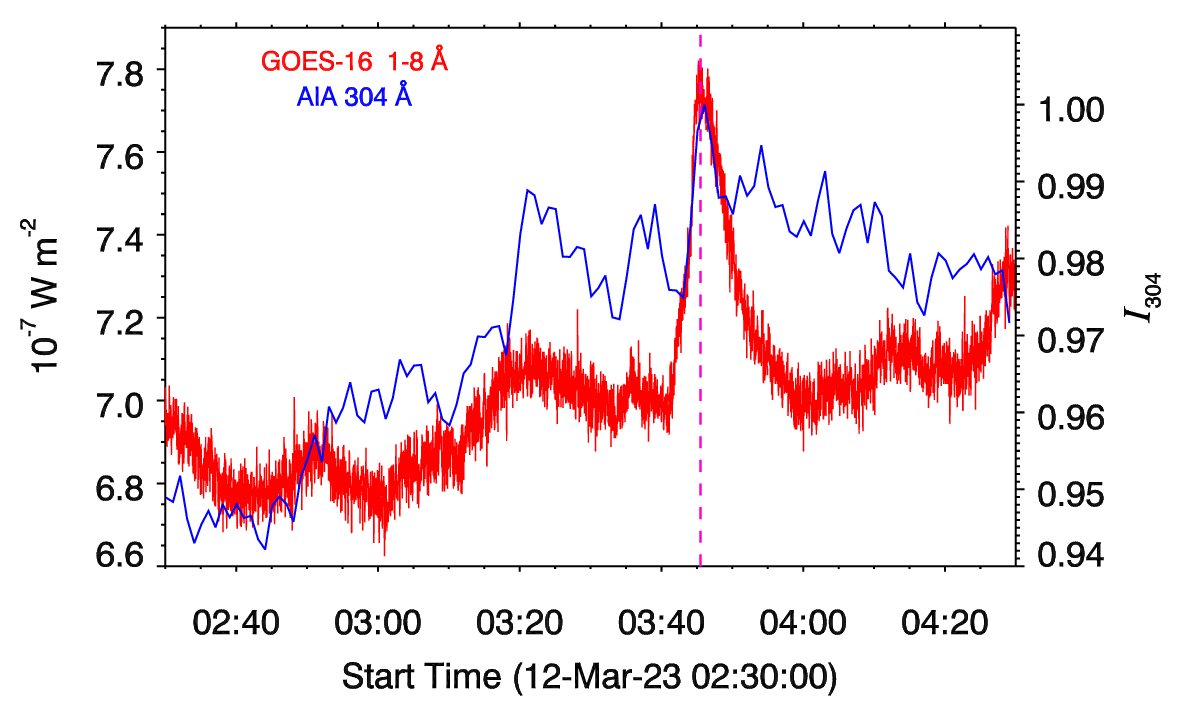}
\centering
\caption{Light curves of the B7.8 class flare in 1$-$8 {\AA} (red line) and 304 {\AA} (blue line) during 02:30$-$04:30 UT.
The magenta dashed line denotes the flare peak time at 03:45 UT.}
\label{fig6}
\end{figure}

The B-class flare occurs close to FP1 as the prominence lifts up. In Figure~\ref{fig2}(a4), a black rectangle is drawn to signify the flare region in EUV wavelengths.
The total intensities within the region during 02:30$-$04:30 UT are calculated and normalized to the maximum.
Figure~\ref{fig6} shows light curves of the flare in 1$-$8 {\AA} (red line) and 304 {\AA} (blue line). The light curves have the same trend and peak time (03:45:30 UT).
It should be emphasized that whether the prominence eruption generates a larger flare with parallel flare ribbons as well as post-flare loops beneath the prominence is unknown 
since the source location is on the farside, which will be presented in Section~\ref{3d}.

\begin{figure}
\includegraphics[width=0.45\textwidth,clip=]{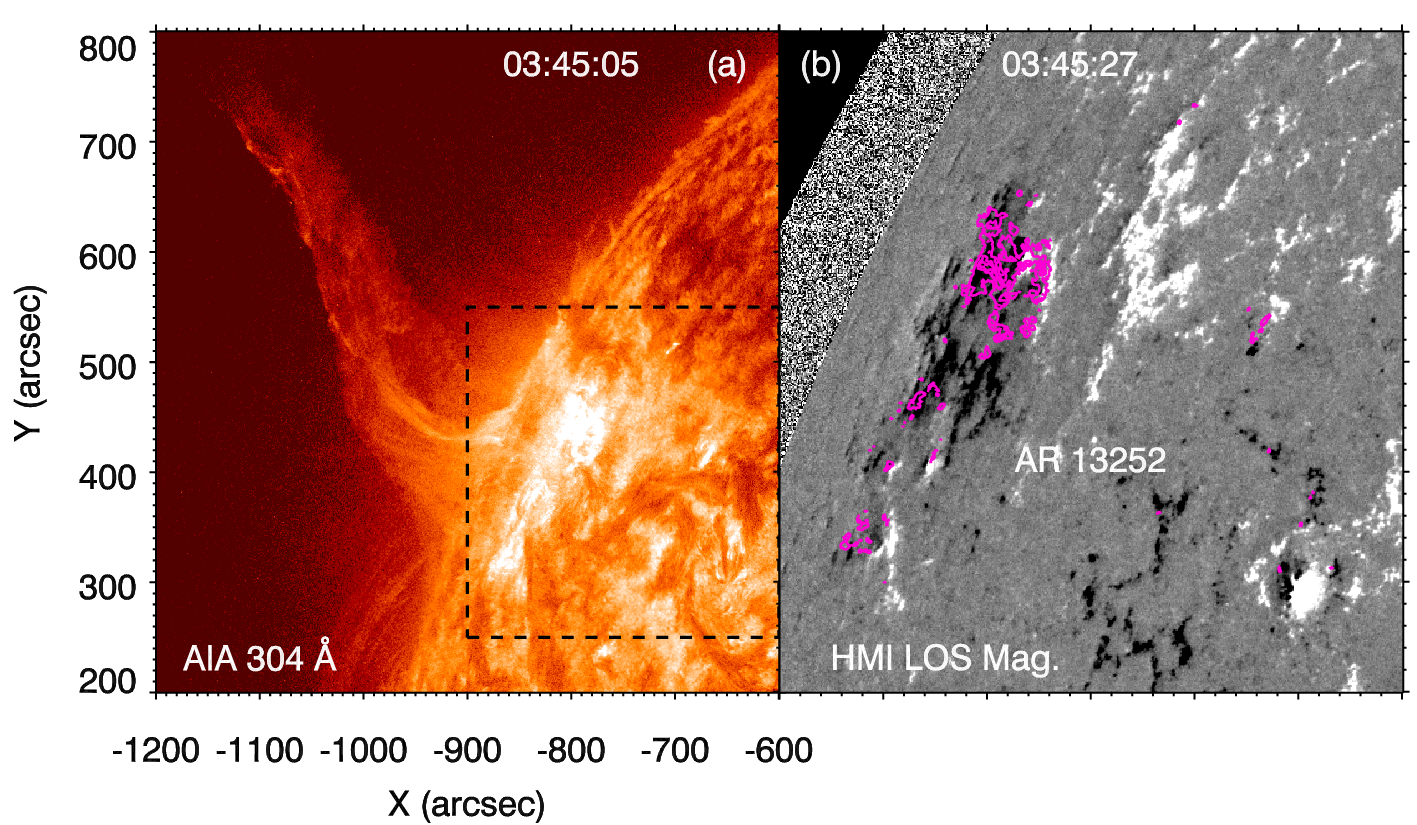}
\centering
\caption{(a) AIA 304 {\AA} image of the prominence and flare at 03:45:05 UT.
The black dashed box denotes the FOV of panel (b).
(b) HMI LOS magnetogram at 03:45:27 UT within the dashed box. 
Intensity contours of the flare region are superposed on AR 13252 with magenta lines.}
\label{fig7}
\end{figure}

In Figure~\ref{fig7}, the left panel shows the AIA 304 {\AA} image of the prominence and flare at 03:45:05 UT.
The black dashed box signifies the FOV of the right panel, which demonstrates the LOS magnetogram of the photosphere observed by SDO/HMI at 03:45:27 UT.
The magenta lines represent 304 {\AA} intensity contours of the flare region superposed on AR 13252, indicating that FP1 of the prominence is rooted in AR 13252.

\begin{figure}
\includegraphics[width=0.45\textwidth,clip=]{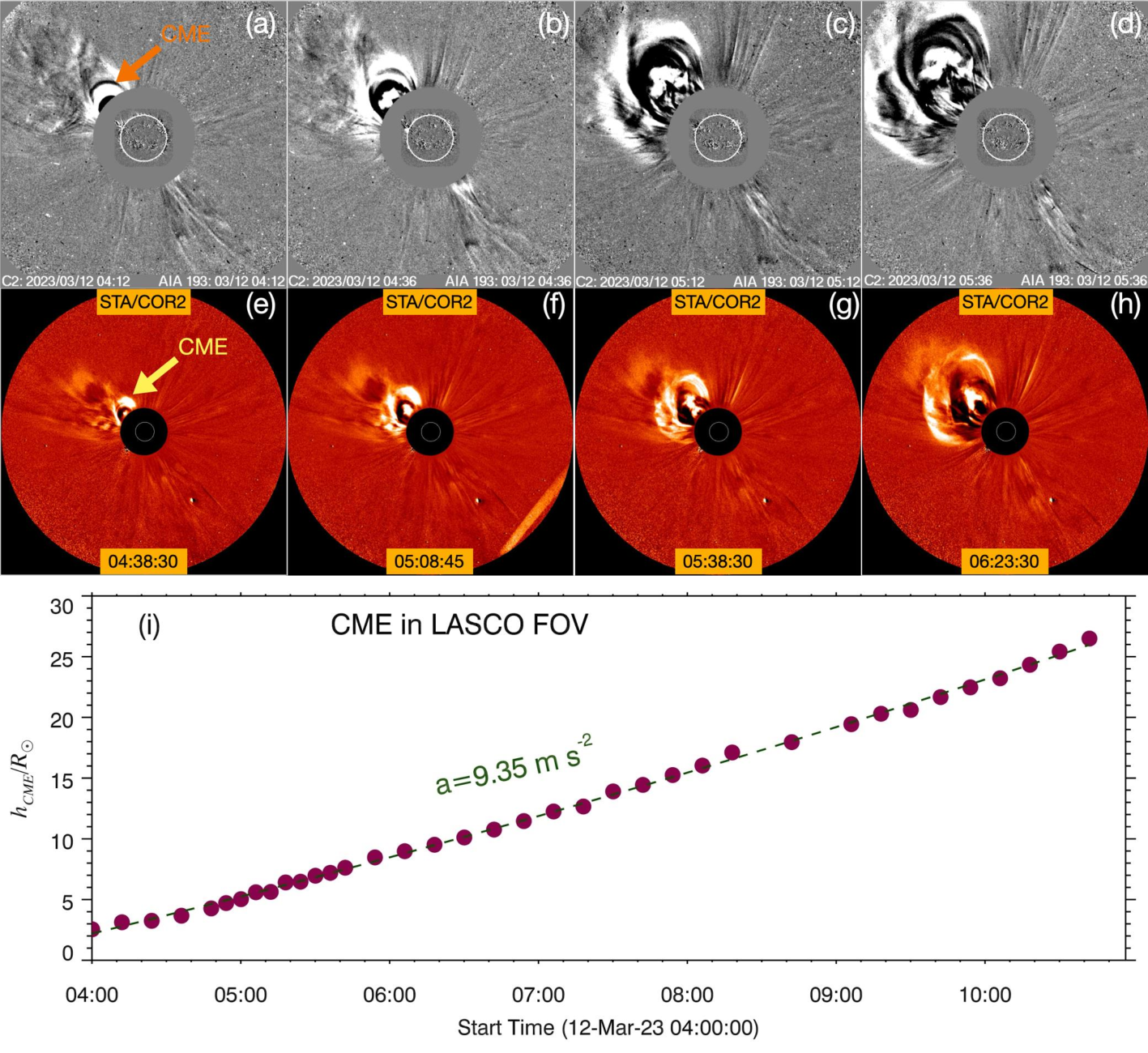}
\centering
\caption{Running-difference images of the CME observed by LASCO-C2 during 04:12$-$05:36 UT (a-d) 
and STA/COR2 during 04:38$-$06:23 UT (e-h). The orange and yellow arrows point to the CME front.
(i) Height-time plot of the CME in SOHO/LASCO FOV. 
The result of a quadratic fitting is plotted with a dark green line, and the CME acceleration is labeled.}
\label{fig8}
\end{figure}

In Figure~\ref{fig8}, panels (a)-(d) show running-difference images of the CME observed by LASCO-C2.
The partial halo CME\footnote{https://cdaw.gsfc.nasa.gov/CME\_list/UNIVERSAL\_ver2/2023\_03/univ2023\_03.html}
first appears at 04:00 UT and propagates in the northeast direction with a central position angle of $\sim$49$\degr$ and an angular width of $\sim$169$\degr$.
Panels (e)-(h) show running-difference images of the CME observed by STA/COR2 during 04:38$-$06:23 UT, 
featuring a typical three-part structure including a bright front, a dark cavity, and a bright, cusp-like core \citep{ill85,song22}.
In panel (i), temporal evolution of the heliocentric distances of the CME front in LASCO FOV during 04:00$-$10:42 UT is drawn with maroon circles.
The whole data are fitted with a quadratic function (dark green line), resulting in a constant acceleration ($\sim$9.35 m s$^{-2}$), 
which is close to the value given by the CDAW CME catalog.
The apparent speed of CME increases from $\sim$573 to $\sim$798 km s$^{-1}$.

\subsection{3D reconstruction and tracking of the prominence} \label{3d}
In the GCS model, the height and half-angular width of the legs are $h$ and $\alpha$, respectively.
$\kappa=\sin\delta$ denotes the aspect ratio of the CME size at two orthogonal directions, where $\delta$ signifies the half angle of a leg.
The total edge-on and face-on angular widths ($\omega_{\mathrm{EO}}$ and $\omega_{\mathrm{FO}}$) are 2$\delta$ and 2$(\delta+\alpha)$, respectively.
The source region of the CME is characterized by the Carrington longitude ($\phi$), latitude ($\theta$), and tilt angle ($\gamma$) of the PIL with respect to the meridian.
Therefore, $\gamma=0\degr$ and $\gamma=90\degr$ stand for MFRs along the meridian and latitude line, respectively.
The height of the CME leading edge is:
\begin{equation} \label{eqn-1}
h_{\mathrm{LE}}=h(\frac{1+\kappa}{1-\kappa^2})(\frac{1+\sin\alpha}{\cos\alpha}).
\end{equation}

\begin{figure*}
\includegraphics[width=0.90\textwidth,clip=]{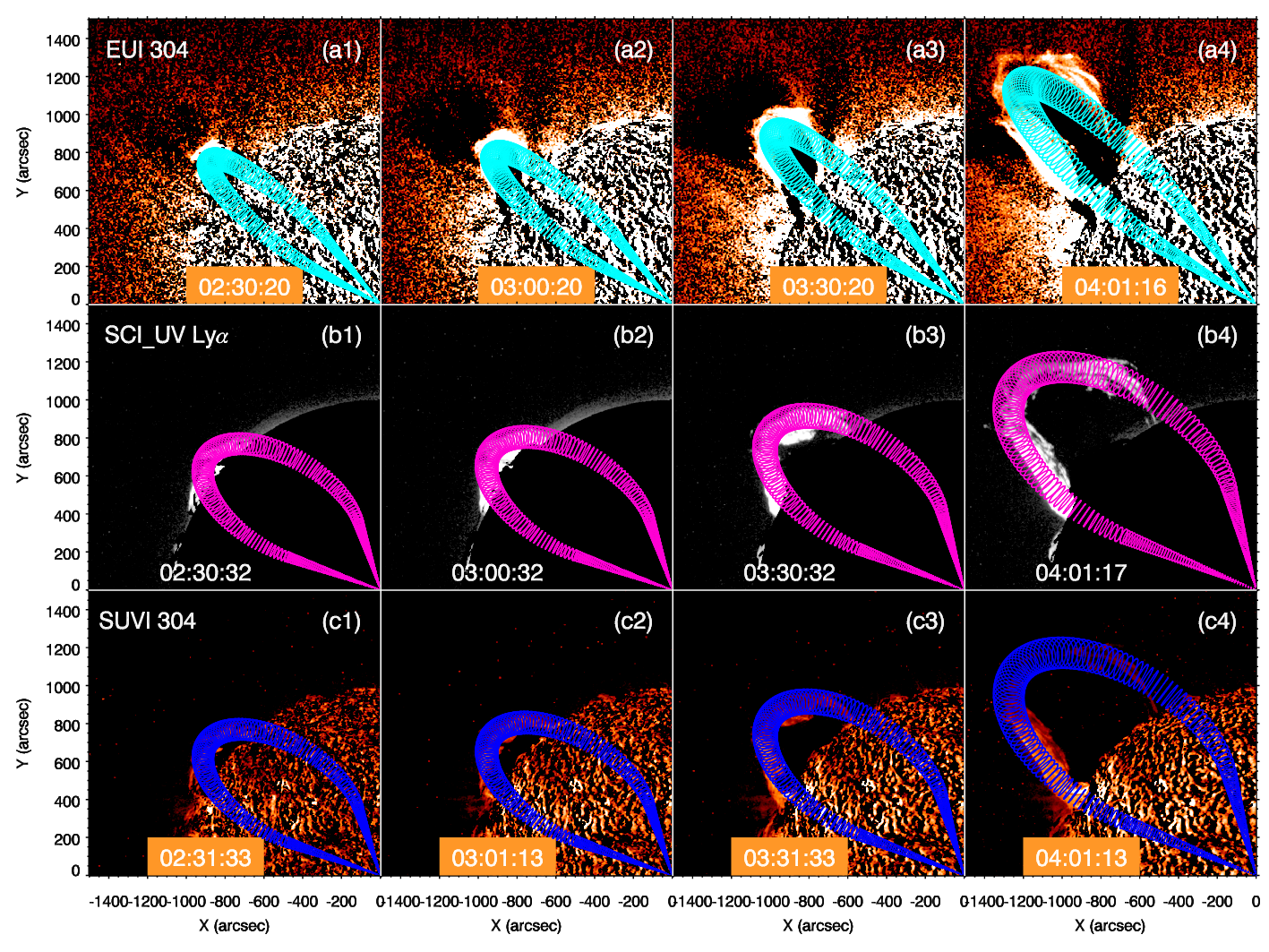}
\centering
\caption{The eruptive prominence observed by EUI (a1-a4), SCI\_UV (b1-b4), and SUVI (c1-c4) during 02:30$-$04:01 UT.
Projections of the reconstructed GCS are superposed on the images with cyan, magenta, and blue lines, respectively.}
\label{fig9}
\end{figure*}

To perform 3D reconstruction of the prominence, we need simultaneous observations from different viewpoints.
Figure~\ref{fig9} shows Ly$\alpha$ images observed by SCI\_UV (b1-b4), base-difference images observed by EUI (a1-a4) and SUVI (c1-c4) during 02:30$-$04:01 UT.
Projections of the reconstructed GCS models are superposed with magenta, cyan, and blue lines. It is clear that the GCS models fit well with the loop-like prominence.
The parameters are listed in Table~\ref{tab-2}. The separation angle (2$\alpha$) between the two legs and aspect ratio ($\kappa$) are $\sim$80$\degr$ and $\sim$0.05.
The total edge-on and face-on angular widths of the prominence are $\sim$6$\degr$ and $\sim$86$\degr$, respectively.
The longitude ($\phi$) is nearly -110$\degr$, meaning that the source region is behind the eastern limb by $\sim$20$\degr$.
The latitude ($\theta$) increases slightly from $\sim$40$\degr$ to $\sim$43$\degr$, which is consistent with the position angle (49$\degr$) of the related CME.
The value of $\gamma$ is equal to $\sim$70$\degr$, suggesting that the PIL associated with the prominence is $\sim$20$\degr$ inclined to the EW direction.
Assuming a radial propagation, the true speed of the CME increases from $\sim$610 to $\sim$849 km s$^{-1}$ during 04:00$-$10:42 UT.
\textbf{The absence of a type II radio burst during the event suggests that the CME likely did not drive a shock wave.}

\begin{deluxetable}{cccccccc}
\tablecaption{Parameters of the GCS modeling during 02:30$-$04:25 UT. \label{tab-2}}
\tablecolumns{8}
\tablenum{2}
\tablewidth{0pt}
\tablehead{
\colhead{$\alpha$} &
\colhead{$\delta$} &
\colhead{$\kappa$} &
\colhead{$\omega_{\mathrm{EO}}$} &
\colhead{$\omega_{\mathrm{FO}}$} &
\colhead{$\phi$} &
\colhead{$\theta$} &
\colhead{$\gamma$}\\
\colhead{(deg)} &
\colhead{(deg)} &
\colhead{} &
\colhead{(deg)} &
\colhead{(deg)} &
\colhead{(deg)} &
\colhead{(deg)} &
\colhead{(deg)}
}
\startdata
40 & 3 & 0.05 & 6 & 86 & -110 & 40$-$43 & 70 \\
\enddata
\end{deluxetable}

\begin{figure}
\includegraphics[width=0.45\textwidth,clip=]{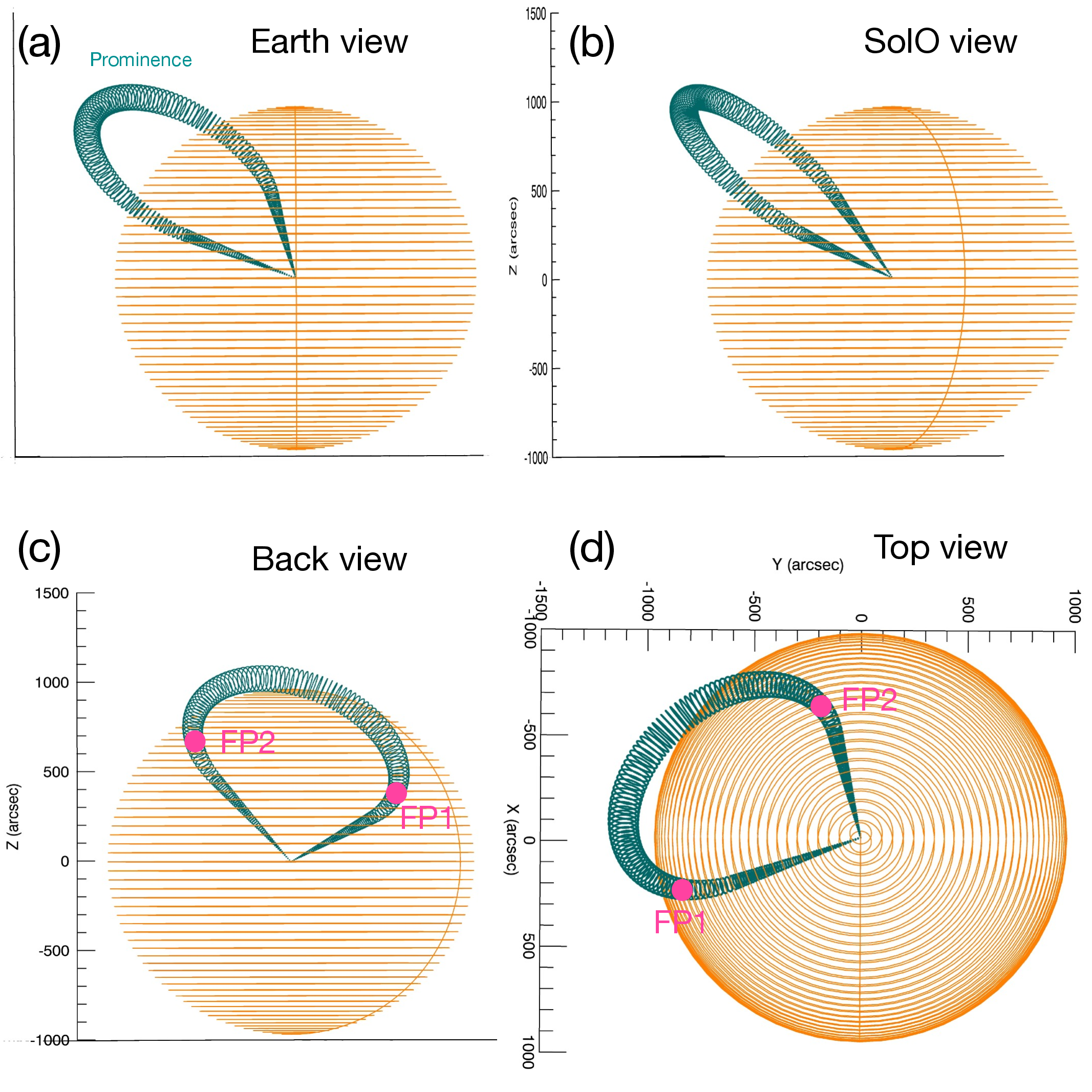}
\centering
\caption{3D visualization of the Sun (orange lines) and reconstructed GCS model (blackish green lines) at 03:45 UT 
from vantage points of Earth (a), SolO (b), farside (c), and solar north pole (d). 
The magenta dots represent the southeast footpoint (FP1) and northwest footpoint (FP2) of the prominence.}
\label{fig10}
\end{figure}

Figure~\ref{fig10} shows 3D visualization of the Sun (orange lines) and reconstructed GCS model (blackish green lines) at 03:45 UT 
from vantage points of Earth (a), SolO (b), farside (c), and solar north pole (d). 
The magenta dots represent the footpoint (FP2) of the prominence on the farside, which is $\sim$896.4 Mm away from the footpoint (FP1) on the frontside.

\begin{figure*}
\includegraphics[width=0.90\textwidth,clip=]{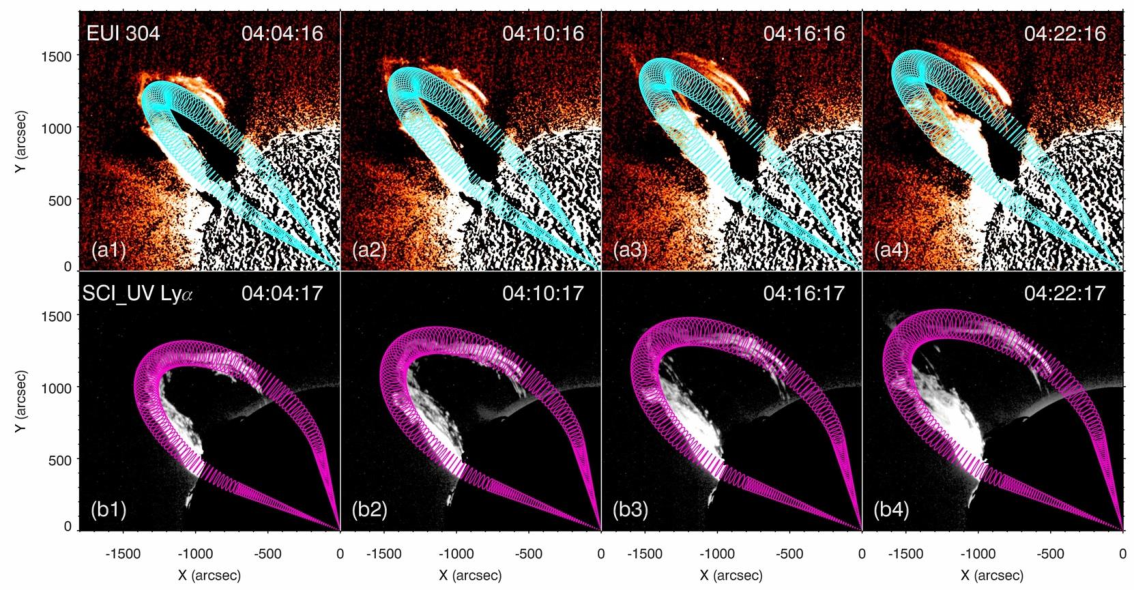}
\centering
\caption{The prominence observed by EUI (a1-a4) and SCI\_UV (b1-b4) during 04:04$-$04:22 UT.
Projections of the reconstructed GCS are superposed on the images with cyan and magenta lines.}
\label{fig11}
\end{figure*}

After 04:01 UT, the top of prominence escapes the GOES-16/SUVI FOV. Consequently, it is impossible to perform 3D reconstruction and tracking using observations from SUVI.
However, we can use simultaneous observations from EUI and SCI\_UV with higher cadences. 
In Figure~\ref{fig11}, the EUI 304 {\AA} and SCI\_UV Ly$\alpha$ images during 04:04$-$04:22 UT are displayed in the top and bottom panels, 
which are superposed by projections of the reconstructed GCS models with cyan and magenta lines.
It is obvious that the modeling is acceptable, despite that the apex of the prominence develops into a cusp structure.

\begin{deluxetable}{cccc}
\tablecaption{Parameters ($h$ and $h_{\mathrm{LE}}$) of the GCS modeling during 02:30$-$04:25 UT. \label{tab-3}}
\tablecolumns{4}
\tablenum{3}
\tablewidth{0pt}
\tablehead{
\colhead{Time} &
\colhead{Instruments} &
\colhead{$h$} &
\colhead{$h_{\mathrm{LE}}$}\\
\colhead{(UT)} &
\colhead{} &
\colhead{(Mm)} &
\colhead{(Mm)}
}
\startdata
02:30 & EUI, SCI\_UV, SUVI & 388.8$\pm$1.8 & 874.8$\pm$4.0 \\
02:45 & EUI, SCI\_UV, SUVI & 396.0$\pm$3.2 & 891.0$\pm$7.3 \\
03:00 & EUI, SCI\_UV, SUVI & 406.8$\pm$4.9 & 915.3$\pm$11.1 \\
03:15 & EUI, SCI\_UV, SUVI & 424.8$\pm$6.2 & 955.8$\pm$14.0 \\
03:30 & EUI, SCI\_UV, SUVI & 446.4$\pm$7.5 & 1004.4$\pm$17.0 \\
03:45 & EUI, SCI\_UV, SUVI & 489.6$\pm$8.7 & 1101.6$\pm$19.7 \\
04:00 & EUI, SCI\_UV & 554.4$\pm$10.8 & 1247.4$\pm$24.3 \\
04:01 & EUI, SCI\_UV, SUVI & 561.6$\pm$11.7 & 1263.6$\pm$26.3 \\
04:04 & EUI, SCI\_UV & 590.4$\pm$13.4 & 1328.4$\pm$30.2 \\
04:07 & EUI, SCI\_UV & 612.0$\pm$14.5 & 1377.0$\pm$32.8 \\
04:10 & EUI, SCI\_UV & 633.6$\pm$16.5 & 1425.6$\pm$37.3 \\
04:13 & EUI, SCI\_UV & 648.0$\pm$17.3 & 1458.0$\pm$39.0 \\
04:16 & EUI, SCI\_UV & 662.4$\pm$19.2 & 1490.4$\pm$43.4 \\
04:19 & EUI, SCI\_UV & 673.2$\pm$20.4 & 1514.7$\pm$46.0 \\
04:22 & EUI, SCI\_UV & 687.6$\pm$21.7 & 1547.1$\pm$49.0 \\
04:25 & EUI, SCI\_UV & 702.0$\pm$23.5 & 1579.5$\pm$53.1 \\
\enddata
\end{deluxetable}

\begin{figure}
\includegraphics[width=0.45\textwidth,clip=]{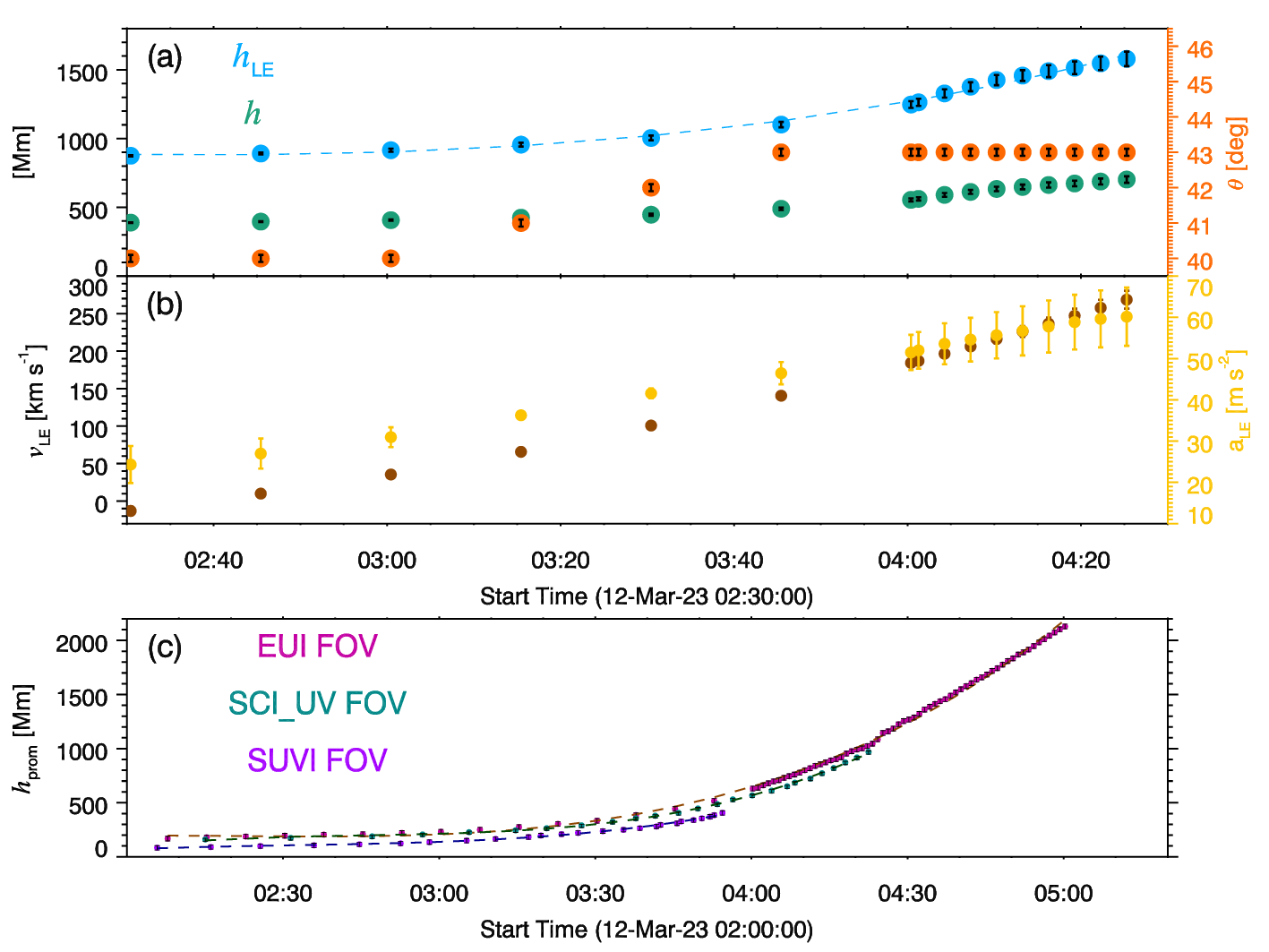}
\centering
\caption{(a)-(b) Temporal evolutions of $h$ (green circles), $h_{\mathrm{LE}}$ (blue circles), $\theta$ (orange circles), 
$v_{\mathrm{LE}}$ (brown circles), and $a_{\mathrm{LE}}$ (yellow circles).
(c) Temporal evolutions of the prominence height above the limb along S1, S2, and S3 in the FOVs of SCI\_UV, SUVI, and EUI, respectively.
Cubic fittings using Equation~\ref{eqn-2} are performed and superposed with dashed lines.}
\label{fig12}
\end{figure}

We carry out GCS modeling for 16 moments between 02:30 UT and 04:25 UT, when the prominence could be distinctly identified.
The values of $h$ and $h_{\mathrm{LE}}$ (Equation~\ref{eqn-1}) are listed in the third and fourth columns of Table~\ref{tab-3} 
and drawn with green and blue circles in Figure~\ref{fig12}(a).
$h$ increases from $\sim$388.8 Mm at 02:30 UT to $\sim$702.0 Mm at 04:25 UT.
The corresponding $h_{\mathrm{LE}}$ increases from $\sim$874.8 Mm (1.26\,$R_{\sun}$) to $\sim$1579.5 Mm (2.27\,$R_{\sun}$).
After 04:30 UT, it is unsuitable to use the GCS modeling for the cusp-like prominence with rapidly decreasing intensities in 304 {\AA} and Ly$\alpha$.

A cubic fitting of $h_{\mathrm{LE}}$ as a function of time ($t$) is conducted:
\begin{equation} \label{eqn-2}
h_{\mathrm{LE}}(t)=c_{0}+c_{1}t+c_{2}t^2+c_{3}t^3.
\end{equation}
The result of curve fitting is drawn with a blue dashed line in Figure~\ref{fig12}(a).
The true speed of the prominence is:
\begin{equation} \label{eqn-3}
v_{\mathrm{LE}}(t)=\frac{dh_{\mathrm{LE}}}{dt}=c_{1}+2c_{2}t+3c_{3}t^2.
\end{equation}
Temporal evolution of $v_{\mathrm{LE}}(t)$ is drawn with brown circles in Figure~\ref{fig12}(b).
$v_{\mathrm{LE}}$ reaches $\sim$269 km s$^{-1}$ at 04:25 UT.
Likewise, temporal evolution of the prominence acceleration ($a_{\mathrm{LE}}(t)=dv_{\mathrm{LE}}/dt$) is drawn with yellow circles in Figure~\ref{fig12}(b), 
which increases from $\sim$24 to $\sim$60 m s$^{-2}$ during 02:30$-$04:25 UT.

In Figure~\ref{fig2}(d3), a long slice (S3) is used to investigate the height evolution of the prominence in EUI FOV.
In Figure~\ref{fig12}(c), trajectories of the prominence above the limb in the FOVs of EUI, SCI\_UV, and SUVI 
are drawn with orchid, light seagreen, and medium orchid dots, respectively.
Thanks to the extraordinarily large EUI FOV, 
the prominence leading edge could be identified up to $\sim$2129 Mm ($\sim$3.06\,$R_{\sun}$) above the limb until 05:00 UT.
Similar cubic fittings for the prominence heights are performed using Equation~\ref{eqn-2} and superposed with dashed lines in Figure~\ref{fig12}(c).
\citet{sch08} studied the early evolutions of two near-limb filament eruptions associated with CMEs. 
It is found that the heights of filaments above the solar limb have a dependence $h(t)\propto t^3$ in their rapid-acceleration phases.

\begin{figure*}
\includegraphics[width=0.90\textwidth,clip=]{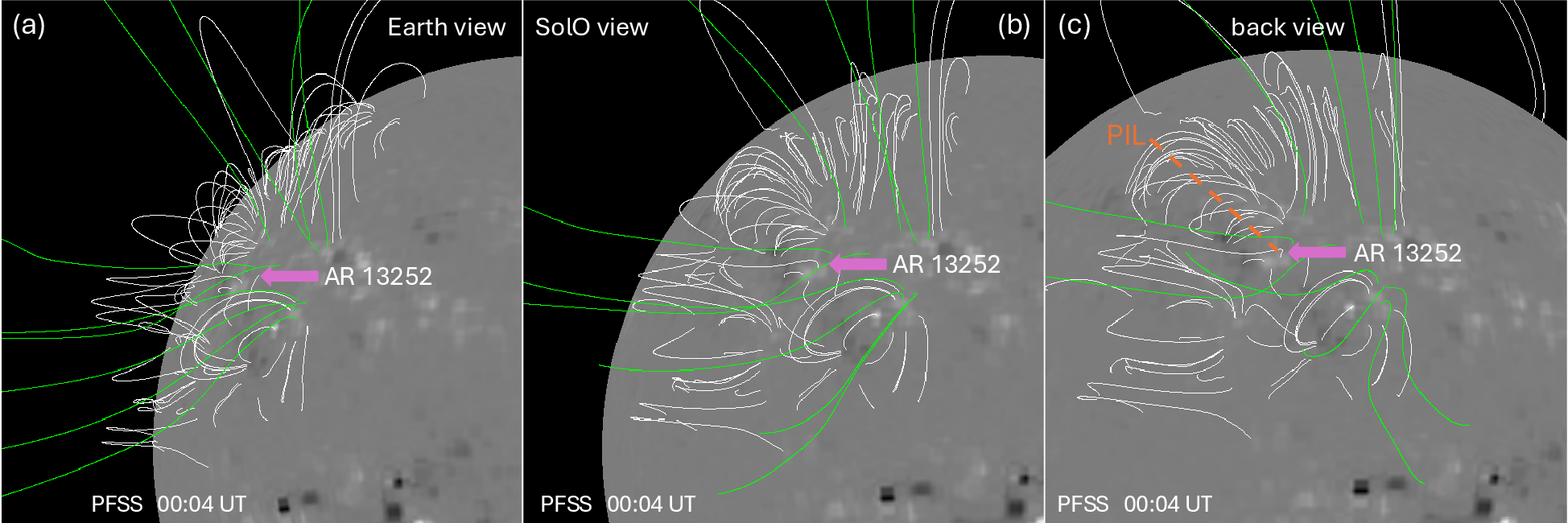}
\centering
\caption{Magnetic field lines at 00:04 UT obtained from the PFSS modeling. The white and green lines represent closed and open field.
The left, middle, and right panels show the magnetic configurations seen from Earth, SolO, and back, respectively.
The pink arrows point to AR 13252, where FP1 of the intermediate prominence takes root in.
In panel (c), the orange dashed line represents the rough positions of PIL associated with the prominence.}
\label{fig13}
\end{figure*}

Figure~\ref{fig13} shows magnetic field lines at 00:04 UT obtained from the potential-field source surface \citep[PFSS;][]{sch69} modeling.
The field lines viewed from Earth, SolO, and back are displayed in the left, middle, and right panels, respectively.
The pink arrows point to AR 13252, where FP1 of the prominence resides. A group of arcade field lines line up to the northeast of AR 13252.
In panel (c), the orange dashed line represents the rough positions of PIL associated with the prominence under the arcade, 
which is generally consistent with the connection between FP1 and FP2 (Figure~\ref{fig10}(c)).
It should be emphasized that since the LOS magnetograms on the farside are unavailable,
precise 3D magnetic configuration above the filament is unknown.

\section{Discussion} \label{dis}
\subsection{Cause of rotation and cusp structure of the prominence} \label{dis-11}
Kink instability takes place when the total twist in a MFR exceeds a critical value \citep{hood81,fan03,kli04,tor04,tor05}. 
Observational evidences of eruptive filaments triggered by kink instability are abundant \citep{ji03,gre07,liu07,guo10,yan14a,yan14b,zhou23}.
Using a flux rope model as the initial condition, \citet{tor05} performed MHD simulations of solar eruptions as a result of helical kink instability.
The results nicely reproduce a confined filament eruption on 2002 May 27 \citep{ji03} and a full eruption on 2001 May 15.
As the flux rope rises up, it experiences a writhing motion. The apex ramps up and develops into a cusp structure.
Meanwhile, the legs show continuous rotations (see their Fig. 1).
\citet{gre07} examined the relationship between magnetic helicity and the direction of rotation of six filaments undergoing kink instability.
It is revealed that sinistral (dextral) filaments with positive (negative) helicity experience clockwise (counterclockwise) rotations.

In the current work, the southeast leg of the prominence experiences rotations in the counterclockwise direction, 
which are clearly demonstrated by spectroscopic observations of CHASE/HIS (Figure~\ref{fig4}) and imaging observations of ASO-S/SCI\_UV (Figure~\ref{fig5}).
Therefore, the prominence could be a dextral filament viewed from the backside before eruption according to the statistical results \citep{gre07}.
The shape of the apex changes from a smooth loop to a cusp structure (Figure~\ref{fig2}) and the northwest leg drifts eastward after 04:30 UT, suggesting a writhing motion.
Hence, the eruption is most likely triggered by ideal kink instability, although the total twist is hard to measure.
Other mechanisms, such as magnetic flux emergence \citep{chen00} and torus instability \citep{kli06}, could not be fully excluded.

\subsection{Implication of the 3D reconstructions} \label{dis-22}
Prominence eruptions and CMEs originating from the farside of the Sun are common. 
Fast CMEs are capable of driving fast-mode shock waves, which further generate solar energetic particles \citep[SEPs;][]{rea13,des16} or
Sustained Gamma-Ray Emission \citep[SGRE;][]{go18}.
\citet{zyj24} studied large-amplitude, vertical prominence oscillations simultaneously observed by AIA and the behind STEREO (hereafter STB) on 2014 September 1.
The oscillation is triggered by an EUV wave associated with a flare and CME originating from AR 12158 on the farside.
\citet{zqm24a} discovered that two successive EUV waves were driven by a fast CME ($\sim$2782 km s$^{-1}$) and a coronal jet ($\sim$831 km s$^{-1}$)
originating from AR 13575 behind the western limb on 2024 February 9. 
The EUV waves excite large-amplitude, transverse oscillations of a quiescent prominence near the solar south pole, 
which finally gets unstable and erupts to from another CME \citep{li25b}.
\citet{zj24} investigated the SEP event at Mars generated by a fast, partial halo CME on 2022 February 15.
Using the GCS modeling and WL images from STA/COR2 and SOHO/LASCO-C2, the source region on the farside is determined.
\citet{mie22} explored the same prominence eruption and the associated CME on that day.
Using the triangulation method, the true heights of three bright knots of the prominence are derived, reaching up to $\sim$6.97\,$R_{\sun}$.
The true velocities are calculated to be $\sim$1700 km s$^{-1}$.
\citet{zhou23} analyzed the eruption of a long prominence originating behind the western limb on 2013 May 13.
The bright, conjugate flare ribbons and post-flare loops as a result of prominence eruption are manifestly detected by STA.
Using the GCS modeling, the source region (W91N15) of the related prominence/CME is derived. 
Additional events originating behind the limb as observed from Earth reported by previous literature are listed in Table~\ref{tab-4}.
Some of them are related to flares, and all of them are associated with shock waves.

In the current study, the separation angles ($<28\degr$) of STA and SolO with the Sun-Earth line are too small to directly detect the source region of the huge, intermediate prominence.
Thanks to the forward modeling of the prominence with the standard GCS model by assuming that the two legs are coplanar in the early stage of evolution, 
we are able to pinpoint both the source region and the second footpoint (Figure~\ref{fig10}).
The connection between the two footpoints is basically in agreement with the direction of a PIL adjacent to AR 13252 (Figure~\ref{fig13}).
Moreover, the true height of the prominence reaches up to $\sim$3.32\,$R_{\sun}$ above the limb and
the true speed (610$-$849 km s$^{-1}$) of the CME is estimated according to the direction of propagation.
It is noted that we did not perform GCS modeling for the CME itself since the separation angle between STA and SOHO is too small.
The longitude of the source region in our case is very close to that of AR 13256 on 2023 March 13 \citep{dre25}.
However, the CME speed is less than half the CME speed in their study.

\begin{deluxetable*}{ccccccc}
\tablecaption{List of events originating from behind the limb as observed from Earth. \label{tab-4}}
\tablecolumns{7}
\tablenum{4}
\tablewidth{0pt}
\tablehead{
\colhead{Date} &
\colhead{Location} &
\colhead{AR} &
\colhead{Flare} &
\colhead{$V_{\mathrm{CME}}$} &
\colhead{Shock} &
\colhead{Reference} \\
\colhead{} &
\colhead{} &
\colhead{} &
\colhead{} &
\colhead{(km s$^{-1}$)} &
\colhead{(Y/N)} &
\colhead{}
}
\startdata
2011/03/21 & W132       &          &           & 1100 & Y &  \citet{rou12} \\
2011/06/04 & W145 &  &  &  & Y & \citet{la13} \\
2011/11/17 &  E110  &  &  &  & Y & \citet{la13} \\
2011/11/03 & E152 &                   & M4.7-X1.4 & 991 & Y & \citet{gom15} \\
2013/08/19 & W171N08 & & & 1149 & Y & \citet{rod21} \\
2014/09/01 & E126N14 & 12158 & X2.4 & 2079 & Y & \citet{plot17}; \citet{zyj24} \\
2014/10/14 & E120S11 & 12192 & M1.1 & 850 & Y & \citet{wit17} \\
2020/11/24 & E163S25 & 12790 &          & 892 & Y & \citet{buc23} \\
2021/07/17 & E140S20 &          & M5     & 1500 & Y & \citet{pes22} \\
2021/12/04 & W149S11 & & & 761 & Y & \citet{chi23} \\
2021/12/20 & W148N02 & & & 391 & Y & \citet{chi23} \\
2022/02/15 & E132N34 & & & 2200 & Y & \citet{pal24}; \citet{zj24} \\
2023/03/13 & E166N17 & 13258 & & 1800 & Y & \citet{dre25} \\
2023/03/13 & E109S12 & 13256 & & 1800 & Y & \citet{dre25} \\
2024/09/09 & E131S13 &          &  X3.3  & 2162 & Y & \citet{go25} \\
\enddata
\end{deluxetable*}

Compared with the triangulation method, forward modeling using the GCS model is more suitable to perform 3D reconstructions 
and tracking of farside prominence eruptions, whose whole structures could not be simultaneously observed from two vantage points.
Besides, the GCS model is superior to the revised cone model and revised GCS model in the reconstructions of quiescent and intermediate prominences 
whose footpoints are far apart.
The only disadvantage of this method, as mentioned in Section~\ref{intro}, is that magnetic field lines supporting the prominence are unknown.
More sophisticated models, such as the FRi3D model \citep{is16} and 3DCORE model \citep{mo18}, should be considered in the future.
As the maximum of the 25$^{th}$ solar cycle approaching, more and more splendid solar eruptions are expected to take place, 
which may have space weather effects not only on Earth, but also on other planets in the solar system \citep{cle09,jak15,pal24}.
A growing number of spacecrafts have been and will be launched to realize multipoint solar observations, 
such as the Polarimeter to UNify the Corona and Heliosphere \citep[PUNCH;][]{def22}
and the Multiview Observatory for Solar Terrestrial Science \citep[MOST;][]{go24}.
Additional case studies using forward modeling are worthwhile to figure out the source regions and interplanetary propagations of spectacular eruptions.
The early evolutions of prominences, including acceleration, expansion, and rotation, may help to get a deeper understanding of the nature of erupting prominences.

\section{Conclusion} \label{con}
In this paper, we carry out multiwavelength and multiview observations of the eruption of an intermediate prominence originating from the farside of the Sun on 2023 March 12.
The southeast footpoint of the prominence is rooted in AR 13252.
The eruption generates a B7.8 class flare and a partial halo CME. The main results are as follows:
\begin{enumerate}
   \item The prominence lifts off at 02:00 UT and undergoes continuing acceleration for nearly three hours. 
   Rotation of the southeast leg of the prominence in the counterclockwise direction is revealed by spectroscopic observations of CHASE/HIS and imaging observations of ASO-S/SCI\_UV.
   The apex of the prominence changes from a smooth loop to a cusp structure during the rising motion and the northwest leg displays a drift motion after 04:30 UT, indicating a writhing motion.
   Hence, the prominence eruption is most likely triggered by ideal kink instability.
   \item For the first time, we apply the GCS modeling in 3D reconstruction and tracking of the prominence for nearly two hours.
   Both the source region (110$\degr$E, 43$\degr$N) and northwest footpoint (162$\degr$E, 44$\degr$N) are pinpointed.
   The edge-on and face-on angular widths of the prominence are $\sim$6$\degr$ and $\sim$86$\degr$, respectively.
   The axis has a tilt angle of $\sim$70$\degr$ with the meridian.
   The heliocentric distance of the prominence leading edge increases from $\sim$1.26\,$R_{\sun}$ to $\sim$2.27\,$R_{\sun}$.
   According to the direction of propagation, the true speed of the related CME increases from $\sim$610 to $\sim$849 km s$^{-1}$. 
\end{enumerate}

\begin{acknowledgments}
We are grateful to the reviewer for valuable comments and suggestions to improve the quality of this article.
The authors appreciate Profs. Weiqun Gan, Haisheng Ji, Hui Li, and Yu Huang in Purple Mountain Observatory for helpful discussions.
SDO is a mission of NASA\rq{}s Living With a Star Program. AIA and HMI data are courtesy of the NASA/SDO science teams.
The CHASE mission is supported by China National Space Administration (CNSA).
The ASO-S is supported by the Strategic Priority Research Program on Space Science, Chinese Academy of Sciences.
Solar Orbiter is a space mission of international collaboration between ESA and NASA, operated by ESA. 
The Extreme Ultraviolet Imager (EUI) is part of the remote sensing instrument package of the ESA/NASA Solar Orbiter mission. 
SUVI was designed and built at Lockheed-Martin\rq{}s Advanced Technology Center in Palo Alto, California.
STEREO/SECCHI data are provided by a consortium of US, UK, Germany, Belgium, and France.
This work is supported by the Strategic Priority Research Program of the Chinese Academy of Sciences, grant No. XDB0560000,
the National Key R\&D Program of China 2021YFA1600500 (2021YFA1600502), 2022YFF0503003 (2022YFF0503000), 
NSFC under the grant numbers 12373065, Natural Science Foundation of Jiangsu Province (BK20231510), 
and Project Supported by the Specialized Research Fund for State Key Laboratory of Solar Activity and Space Weather.
\end{acknowledgments}

\end{document}